\documentclass[a4paper,11pt]{article}
\usepackage{pos}

\title{Discrete symmetry and 't Hooft anomalies for 3450 model}

\author*{Tetsuya Onogi}
\author{Hiroki Wada}
\author{Tatsuya Yamaoka}

\affiliation{Department of Physics, Osaka University,\\
 1-1 Machkaneyama-cho, Toyonaka, 560-0043, Japan}

\emailAdd{onogi@het.phys.sci.osaka-u.ac.jp}
\emailAdd{t\_yamaoka@het.phys.sci.osaka-u.ac.jp}
\emailAdd{hwada@het.phys.sci.osaka-u.ac.jp}

\abstract{We report our study of the discrete symmetry for lattice 3450 model  proposed by Wang and Wen~\cite{Wang:2013yta} .  Lattice 3450 model is expected to describe the anomaly free chiral U(1) gauge theory in 1+1 dimension using 2+1 dimensional domain-wall fermion with gapping interactions for the mirror sector. We find that the lattice model has exact discrete symmetry in addition to U(1) x U(1) symmetry. Assuming the Zumino-Stora procedure works also for discrete symmetry, we compute the full 't Hooft anomaly for the target continuum U(1) chiral gauge theory with the same discrete symmetry. We show that the mixed and self anomalies involving the discrete symmetry are absent,  which is consistent with the expectation that the lattice 3450 produces chiral U(1) gauge theory in the continuum limit.
\\

OU-HET-1261
}

\FullConference{The 41st International Symposium on Lattice Field Theory (LATTICE2024)\\
 28 July - 3 August 2024\\
Liverpool, UK\\}

\begin{document}
\maketitle

\section{Introduction}
Many people would agree that chiral gauge theory on the lattice is needed. This is because standard model is chiral,  and also,  interesting dynamics in chiral gauge theories are not fully understood yet. 

What about the status of lattice formulation?
There are two approaches. 
One approach is with overlap fermion which has given partial success. 
Abelian gauge theory in constructed by L\"{u}scher~\cite{Luscher:1998du}.
For non-abelian case, it is not completed due to the lattice cohomology problem~\cite{Luscher:2000hn}.
The other approach is Domain-wall fermion~\cite{Kaplan:1992bt} with symmetric gapping interactions~\cite{Fidkowski:2009dba}, \cite{Wang:2022ucy}. 
With gapping interaction, one hopes to gap out mirror edge mode to construct chiral gauge theory  on the lattice. 
The lattice 3-4-5-0 model is one of the examples. And this is the topic I would like to focus in our study.

Here we  would like to make two remarks.  First remark is that for a given interaction, whether mirror  edge mode  can really be gapped out or not is still unknown and numerical test is needed.
Second is that 
However there is a hint that the existence of edge mode in domain-wall fermion is guaranteed by anomaly inflow.
Therefore, symmetry and anomaly give strong restrictions on the symmetric gapping scenario with DW fermion.

The question we would like to ask is 
whether symmetric gapping scenario is consistent with anomaly inflow.
In 3450 model, this is already considered for continuous U(1) symmetries.
There is a study on the  anomaly using cobordism \cite{Wan:2018bns} where it was shown that the (1+1)D nonperturbative 
global anomalies are classified by cobordism group which turn out to vanish. 
In Ref.\cite{Wang:2022fzc}, the discrete symmetry of 1+1d 3450 model is studied and explains how to avoid the CT or P problem (like strong CP) can be avoided. It is not yet known whether the anomalies for possible discrete symmetries such kind also vanish or not. However, there may be a possibility that the model may contain other discrete symmetries which are not included in the continuous $U(1)$ symmetry. If they exist it is worth knwoing whether the anomalies for possible discrete symmetries such kind also vanish or not. 
Therefore, our goal is to study the 3450 model with single flavor as well as two flavors to see 1) whether there are discrete symmetries which are not included in continuous U(1) symmetries other than those in Ref.~\cite{Wang:2022fzc}, and 2)  whether the bulk anomaly including such discrete symmetries. These  would give a new consistency check of the mirror edge mode gapping scenario for lattice 3450 model.

\section{Symmetric gapping and domain-wall fermion}
We will now give a brief review of symmetric gapping with DW fermion.
How can we define chiral gauge theory on the lattice?
The idea is to gap out mirror edge mode from DW fermion 
by symmetric interactions $V_{\rm int}$ at the mirror edge.
There are several promising proposals
 such as SO(10), SM in 4 dim or 3450 model in 2-dim with symmetric gapping 
 \cite{Chen:2012di, Wang:2013yta, You:2014oaa, Kikukawa:2017gvk, Kikukawa:2017ngf,  Wang:2018ugf, Zeng:2022grc, Lu:2022qtc, Lu:2023mlc, Xu:2025hfs}

And numerical test of mirror edge mode decoupling has been carried out.

We would like to consider what conditions are needed for gapping interactions.
 Free DW fermion has exact vector symmetries G.
 Interaction Vint breaks G to a subgroup H.
There are two  necessary conditions for H and $V_{\rm int}$.
\begin{enumerate}
\item  Bulk anomaly for symmetry H must be absent.\\
     Otherwise, anomaly inflow requires massless edge-mode.
     This gives an obstruction against gapping of mirror fermion edge mode.
\item Instanton saturation must be fulfilled by $V_{\rm int}$.\\
    In U(1) instanton sector, there are chiral zero mode at the mirror edge. 
    In order to have nonzero path-integral Vint must saturate mirror zero modes.
\end{enumerate}

\section{3450 model}
3450 model in the continuum is a chiral U(1) gauge theory in 1+1 dimensions with 4 Weyl fermions. 
The chiralities are Left-, Left-, Right-, Right, and the charges are 3, 4, 5, 0.
The action for both dynamical U(1) gauge field and 4-Weyl fermions is given as
\begin{eqnarray}
S =\int dt dx 
[ -\frac{1}{4} F_{\mu\nu} F^{\mu\nu}
 + \sum_{Q=3,4}\bar{\psi}^Q_L \gamma^\mu D_\mu^Q\psi^Q_L + \sum_{Q=5,0}\bar{\psi}^Q_R \gamma^\mu D_\mu^Q\psi^Q_R], 
\end{eqnarray}
where $D_\mu$ is the covariant derivative with charge Q gauge interaction.
One can see that the theory is free from U(1) gauge anomaly and gravitational anomaly.

\section{Domain-wall realization}
Corresponding to the continuum theory, we consider 4 species of DW fermions with U(1) charge (3,4,5,0) with Left-, Left-, Right-,Right- Weyl fermions on the physical edge. 
$V_{\rm int}$ are given by two 6-fermion interactions given here.
\begin{eqnarray}
V_{\rm int} \sim g_1 V_1
               + g_2 V_2 + h.c, 
\end{eqnarray}
where
\begin{eqnarray}
V_1=\psi^3 \bar{\psi^4}(\partial_x \bar{\psi^4})\psi^5 \psi^0(\partial_x\psi^0), ~~~
V_2=\psi^3(\partial_x\psi^3)\psi^4 \bar{\psi}^5 (\partial_x\bar{\psi}^5)\psi^0
\end{eqnarray}
As we will see in a moment, it is designed to satisfy the necessary conditions.
Assuming that this $V_{\rm int}$ is the correct choice for the moment, let us consider the symmetry H. 
One finds that with interaction $V_{\rm int}$, $G=U(1)^4$ symmetry symmetry is broken down to $H=U(1) \times U(1)^\prime$. 
$U(1)$ and $U(1)^\prime$ have charges given 
\begin{eqnarray}
q=(-1,2,1,2) , ~~ q^\prime=(2,1,2,-1)
\end{eqnarray}
 and linear combination of the two U(1) contains U(1) charge for the dynamical gauge field, which is $q_{em}=(3,4,5,0)=q+2q^\prime$.

In this model, the gapping interaction and symmetry H satisfy the necessary condition as shown here. 
\begin{enumerate}
\item Bulk anomaly is zero. \\
The bulk anomalies for $U(1)$, $U(1)^\prime$  are zero including mixed anomalies.
\begin{eqnarray}
q_3^2 + q_4^2 -q_5^2 -q_0^2 =0\\
q_3^{\prime 2} + q_4^{\prime 2} -q_5^{\prime 2} -q_0^{\prime 2} =0\\
q_3 q_3^{\prime} + q_4 q_4^{\prime} -q_5 q_5^{\prime} -q_0 q_0^{\prime} =0
\end{eqnarray}
\item Instanton saturation\\
And combining $V_1$ and $V_2$ one can generate ‘t Hooft vertex which makes instanton saturation.
\begin{eqnarray}
V_1^\dagger(V_2)^2 &\sim &\psi^3\partial_x(\psi^3)\partial^2_x(\psi^3)
\nonumber\\
&&\times~ \psi_4\partial_x(\psi_4)\partial^2_x(\psi_4)\partial^3_x(\psi_4)
\nonumber\\
&&\times~ \psi^\dagger_5\partial_x(\psi^\dagger_5)
\partial_x^2(\psi^\dagger_5)\partial^3_x(\psi^\dagger_5)
\partial_x^4(\psi^\dagger_5)
\end{eqnarray}
\end{enumerate}

Moreover, both semi-classical analysis after bosonization and discussion based on FQHE shows that symmetric gapping does take place with the gapping interaction

\section{Discrete symmetries in 3450 model}
We now study what is the discrete symmetry allowed by $V_{\rm int}$  in 3-4-5-0 mode l for single flavor.
Under general  U(1) transformation for the fermion parametrized by $\alpha_Q$, 
\begin{eqnarray}
\psi^Q \rightarrow \exp(i \alpha_Q) \psi^Q  ~~~(Q=3,4,5,0)
\end{eqnarray}
$V_{\rm int}$  transforms as 
\begin{eqnarray}
V_1 \rightarrow \exp(i(\alpha_3 -2\alpha_4+\alpha_5+2\alpha_0)) V_1 \\
V_2 \rightarrow \exp(i(2\alpha_3 +\alpha_4-2\alpha_5+\alpha_0)) V_2
\end{eqnarray}
Therefore, invariance of $V_{\rm int}$ requires the condition:
\begin{eqnarray}
\alpha_3 -2\alpha_4+\alpha_5+2\alpha_0=2\pi n, \\
2\alpha_3 +\alpha_4-2\alpha_5+\alpha_0=2\pi m
\end{eqnarray}
Here n, m are arbitrary integers.
Using $U(1)\times U(1)^\prime$ transformation, you can make two of the $\alpha$’s vanish.
This gives you the solutions given as
\begin{eqnarray}
\alpha^{(1)}=\frac{2\pi}{3} (1, 0,0,1),  &
\alpha^{(2)}=\frac{2\pi}{3} (0, 1,-1,0), &
\alpha^{(3)}=\frac{2\pi}{3} (0, 1,0,-1)
\\
\alpha^{(4)}=\frac{2\pi}{3} (1,0, -1,0), &
\alpha^{(5)}=\frac{2\pi}{3} (0,0,1,2), &
\alpha^{(6)}=\frac{2\pi}{3} (2, 1,0,0)
\end{eqnarray}
For example,  the first solution give $Z_3$ symmetry 
\begin{eqnarray}
\mbox{e.g.}~~ \alpha^{(1)}:& ~\psi_3\rightarrow \omega_3\psi_3, ~~\psi_0\rightarrow \omega_3 \psi_0\\
&~\psi_{4,5} ~~\mbox{invariant}.  ~~~~~~(\omega_3 =\exp(2\pi i/3))
\end{eqnarray}
However, it turns out that these discrete symmetries are included in contiinuous U(1)x U(1)  symmetry. 
\begin{eqnarray}
\mbox{e.g.}~\frac{2\pi}{3}(q+q^\prime)=\frac{2\pi}{3}(1,3,3,1) = \alpha^{(1)} ~(\mbox{mod }2\pi \mathbb{Z})
\end{eqnarray}
So there is no 0-form discrete symmetry for a single flavor.
\subsection{Discrete symmetry for multi-flavor case}
Next we study the discrete symmetry for multi-flavor case.
In this case $V_{\rm int}$ can be given by this. 
Then the symmetry reduces to $U(1)\times U(1)\times S_{NF}$, where$ S_{NF}$ is the permutation group.
Therefore multi-flavor system has 0-form discrete symmetry.

\section{'t Hooft anomaly}
Let us now compute 't Hooft anomaly. 
For simplicity let us consider 2-flavor case.
The symmetry is $U(1) \times U(1) \times S_2$. 
After gauging these symmetries, we have  three  gauge fields :

One is dynamical U(1) gauge field denoted by $a$. 
Another is U(1) 't Hooft gauge field denoted by $A$.
The last one is the $Z_2$ gauge field, since $S_2$ is isomorphic to $Z_2$. 
This is described by a set of 1-form and 0-form gauge fields B1 and B0  satisfying this constraint.
\begin{eqnarray}
2B^{(1)}= dB^{(0)}
\end{eqnarray}
And the field satisfies
\begin{eqnarray}
\frac{1}{2\pi}\int_{\Sigma_1} dB^{(0)} \in \mathbb{Z} \longrightarrow
\frac{1}{2\pi}\int_{\Sigma_1} B^{(1)} \in \frac{1}{2}\mathbb{Z}
\end{eqnarray}
Denoting the two flavor fermion with charge $Q$ as
$\psi_Q=\left(
\begin{array}{c}
\psi_Q^1\\
\psi_Q^2
\end{array}
\right)$, 
the symmetry transformation is given as
\begin{eqnarray}
U(1)_Q\times U(1)q ~&~: ~\psi_Q\rightarrow\exp(\i Q\theta_Q+ i q_Q \theta_q) \psi_Q\\
S_2 ~&~:~ \psi_Q\rightarrow
\left(
\begin{array}{cc}
0 & 1\\
1 &0
\end{array}
\right)
\psi_Q\\
\end{eqnarray}
By changing the basis as
$\tilde{\psi}_Q 
=
\frac{1}{\sqrt{2}}
\left(
\begin{array}{cc}
1 & 1\\
1 & -1
\end{array}
\right)
\psi
$, 
the transformation becomes
\begin{eqnarray}
U(1)_Q\times U(1)q &: \tilde{\psi}_Q\rightarrow\exp(i Q\theta_Q+ i q_Q \theta_q) \tilde{\psi}_Q\\
S_2 &: \tilde{\psi}_Q\rightarrow
\left(
\begin{array}{cc}
1 & 0\\
0 & -1
\end{array}
\right)
\tilde{\psi}_Q\\
\end{eqnarray}
so that only the lower component transforms non-trivially under $Z_2$.
We employ the Stora-Zumino procedure and compute the 't Hooft anomaly
4 dimensional SPT action is given as 
\begin{eqnarray}
S_{SPT}
= \frac{2\pi}{2! (2\pi)^2}
\int_{M_4} \left(\sum_{Q=3,4} tr(\mathcal{F}_Q^2 ) - \sum_{Q=5,0} tr(\mathcal{F}_Q^2 ) \right)
\end{eqnarray}
where 
\begin{eqnarray}
\mathcal{F}_Q
= 
\left(
\begin{array}{cc}
Q f+ q_Q F &0\\
0 & Q f + q_Q F +dB^{(1)}
\end{array}
\right)
\end{eqnarray}
Using the discent equation 
the 3-dimensional topological action is 
\begin{eqnarray}
S_{3dim} = \frac{2\pi}{2!(2\pi)^2} \int_{M_3} 
4B^{(1)}(f+ F)
\end{eqnarray}
where anomalies which do not involve discrete gauge field  and only involving continuous $U(1)$ symmetries vanish as already shown before. 
From this we can obtain the mixed anomaly involving the discrete symmetry as
\begin{eqnarray}
\mathcal{A} &= \frac{2\pi}{2! (2\pi)^2} \int_{M_2} 4 \frac{2\pi}{2} (f+F)
&=2\pi \frac{1}{2\pi} \int_{M_2} (f+F) \in 2\pi \mathbb{Z}
\end{eqnarray}
Thus we  find  that 't Hooft anomalies involving discrete symmetry 
cancel for mixed-anomaly,
formally cancel for self-anomaly (except for mathematical subtleties).
This gives further consistency check from the t' Hooft anomaly for discrete symmetry that the symmetric gapping with $V_{\rm int}$ can work. A more mathematically rigorous analysis is in progress.

\section{Summary}
Lattice 3450 model is expected to realize 3450 chiral gauge theory in the continuum by symmetric gapping.
If bulk anomaly exist, it contradicts with symmetric gapping because anomaly inflow requires edge modes. 
We discovered that there is a discrete symmetry for lattice 3450 model with multiple flavors.
We examined 't Hooft anomalies including discrete symmetry via Stora-Zumino procedure and found that they seem to vanish. 
This gives further evidence that symmetric gapping  works for lattice 3450 model. 
We would like to apply this kind of analysis to other interesting models.

\acknowledgments
The work of H.W. was supported in part by JSPS KAKENHI Grant-in-Aid for JSPS fellows Grant Number 24KJ1603. The work of T.O. was supported in part by JSPS KAKENHI Grant Number 23K03387. The work of T.Y. was supported in part by JST SPRING, Grant Number
JPMJSP2138.
\bibliographystyle{JHEP}
\bibliography{Refs}

\end{document}